\documentstyle[12pt,aaspp4,epsfig,flushrt,natbib]{article}
\begin{document}

\title{Discovery of a Second High Frequency QPO from the Microquasar 
GRS 1915+105}
\author{Tod E. Strohmayer}
\affil{Laboratory for High Energy Astrophysics, NASA's Goddard Space Flight 
Center, Greenbelt, MD 20771; stroh@clarence.gsfc.nasa.gov}

\begin{abstract}

We report the discovery in archival Rossi X-ray Timing Explorer (RXTE) data of 
a $\sim$ 40 Hz quasiperiodic oscillation (QPO) in the hard X-ray flux from the 
galactic microquasar GRS 1915+105. The QPO is detected only in the hard X-ray 
band above $\sim 13$ keV and was discovered in observations in which the
previously known 67 Hz QPO is present. The 40 Hz QPO has a typical rms 
amplitude of $\sim 2 \%$ in the 13 - 27 keV band, and a width of $\sim 8.5$ Hz
(FWHM). We show that the 67 and 40 Hz QPOs are detected in the same 
observations in July and November, 1997. However, the QPO is not detected in 
observations from April, May and June, 1996 in which the 67 Hz QPO was first
discovered. The frequency of the 67 Hz QPO is significantly higher in the 1997 
observations by about $5 \%$ compared with the 1996 data. 
The identification of the 40 Hz QPO makes GRS 1915+105 the second black hole 
binary to show a pair of simultaneous high frequency QPO (the other being GRO 
J1655-40). The similarities between the properties of the 67 Hz QPO in GRS 
1915+105 and the recently discovered 450 Hz QPO in GRO J1655-40 suggest that 
the pairs of frequencies in these systems may be produced by the same physical 
mechanism, with the frequency differences between the two being likely due to 
different black hole masses in the two systems. We discuss the implications 
of our result for the mass and spin of GRS 1915+105 as well as for models of 
X-ray variability in black holes and neutron stars.

\end{abstract}


\centerline{Accepted for Publication in Astrophysical Journal Letters}

\vfill\eject

\section{Introduction}

In the last several years quasiperiodic X-ray brightness oscillations (QPOs) 
have been discovered with the Rossi X-ray Timing Explorer (RXTE) in more than 
20 neutron star low mass X-ray binaries (see van der Klis 2000 for a recent 
review). The phenomenology of these QPO is complex, but a key feature is that
pairs of QPO are observed. X-ray QPO in black hole binaries have also been 
observed with RXTE. For example, Remillard et al. (1999) reported the presence 
of a weak, broad $\sim 300$ Hz QPO at times when the X-ray spectrum was hardest
(dominated by a power law component) and the luminosity above $\sim 0.2 
L_{Edd}$. Recently, Strohmayer (2001) reported the discovery of a second QPO,
at 450 Hz, in the hard X-ray flux from GRO J1655-40. This demonstrated for the
first time that black hole systems can also show a pair of high frequency QPO
simultaneously, and suggests that the black hole in GRO J1655-40 has 
appreciable angular momentum. High frequency QPOs have also been observed in 
three black hole transients; 4U 1630-47 at 184 Hz, XTE J1858+226 at 170 Hz, and
XTE J1550-564 between 100 - 283 Hz (see Remillard \& Morgan 1998; Markwardt, 
Swank, \& Taam 1999; Remillard et al. 1999a; Homan et al 2000). 

The high frequencies of these X-ray QPO suggest they are produced in the 
innermost region of the accretion flow near the black hole event horizon. 
When a detailed understanding of the processes involved in their formation 
emerges they should provide a wealth of information on black hole mass and spin
as well as the structure of strongly curved spacetime. A number of different 
models have been proposed to explain these oscillations, but all require that 
strong-field GR effects be taken into account (see Milsom \& Taam 1997; Nowak 
et al. 1997; Wagoner 1998; Stella, Vietri \& Morsink 1999; Merloni et al. 
1999). 

A stable $\sim 67$ Hz QPO has also been observed from the Galactic microquasar 
GRS 1915+105 (Morgan, Remillard \& Greiner 1997). GRS 1915+105 exhibits a broad
range of correlated multiwavelength variability, including very complex 
X-ray behavior (see for example, Muno, Morgan \& Remillard 1999; Markwardt, 
Swank \& Taam 1999; Lin et al. 2000), superluminal radio outflows (Mirabel 
\& Rodriquez 1994; Fender \& Pooley 2000), and correlated Infrared, X-ray and 
radio variability (Fender et al. 1997; Eikenberry et al. 1998; Bandyopadhyay
et al. 1998). Motivated by the discovery of a 2nd high frequency QPO in GRO 
J1655-40 we analysed the archival RXTE data from GRS 1915+105 to search for a 
2nd QPO in this object as well. In this Letter we report the discovery of a 
$\sim 40$ Hz in the hard X-ray flux from GRS 1915+105. We show that this new 
QPO is sometimes present at the same time as the previously discovered 67 Hz 
QPO (Morgan, Remillard \& Greiner 1997), however it is not always seen when the
67 Hz QPO is detected. This is now the 2nd example of a black hole binary with
evidence for a {\it pair} of high frequency QPO and suggests that this 
phenomenology may be common in black hole systems as well as the neutron star
binaries. We discuss briefly the implications of our findings for models of 
X-ray variability in neutron star and black holes systems as well as the 
implications for the mass and spin of the black hole in GRS 1915+105.

\section{Data Analysis}

GRS 1915+105 has been the subject of extensive sets of observations with 
RXTE, both public and proprietary. We began a reanalysis of some of the 
archival RXTE GRS 1915+105 data motivated by the discovery in GRO J1655-40 of 
a 450 Hz QPO in the X-ray band above 13 keV (Strohmayer 2001). Since a second
QPO was found in GRO J1655-40 during observations in which a previously
known 300 Hz QPO was seen, we essentially asked the same question with regard 
to GRS 1915+105; might a second QPO also be seen in the hard X-ray flux in 
observations in which the known 67 Hz QPO is detected? To address this question
we began by analysing the RXTE/PCA AO2 public observations of GRS 1915+105 
under proposal 20402. A total of 97 observations were carried out under this 
program and the data modes included a high time resolution event mode (sampling
rate of 16,384 Hz) covering the $> 13$ keV energy band. We first computed an 
average power spectrum for each observation separately using just the $> 13$ 
keV data. We computed power spectra using 64 s intervals and 0.0625 Hz 
frequency resolution in the range from 0.0625 - 1024.0 Hz. We detected the 67 
Hz QPO in the 5 observations listed in Table 1. A QPO feature near $\sim 40$ Hz
is apparent by visual inspection in some of the average power spectra of these 
5 observations. To search most sensitively for this second feature we computed 
an average power spectrum of all 5 observations. This power spectrum is shown 
in Figure 1. The QPO feature at 40 Hz can be clearly seen in this average 
power spectrum. 

\begin{table}[htbp]
\caption{RXTE AO2 Observations of GRS 1915+105 with 67 Hz QPO}

\begin{tabular}{ccc}

\tableline
Obsid & Epoch (UTC mm/dd/yy) & $\nu_0$ (Hz) \cr
\tableline
20402-01-38-00 & 07/20/97 & $69.32 \pm 1.04$ \cr
20402-01-39-00 & 07/25/97 & $69.09 \pm 0.16$ \cr
20402-01-39-02 & 07/29/97 & $68.53 \pm 0.59$ \cr
20402-01-55-00 & 11/17/97 & $69.01 \pm 0.21$ \cr
20402-01-56-00 & 11/22/97 & $68.76 \pm 0.15$ \cr
\tableline
\end{tabular}

\end{table}

To estimate the significance of this feature we first rescaled 
the power spectrum so that the local mean near 40 Hz was 2 (the value 
expected for a purely Poisson process), and then computed the probability of 
obtaining a power $P = P_{max}\times 510\times 128$ from a $\chi^2$ 
distribution with $2\times 510\times 128$ degrees of freedom. 
Here $P_{max}$ is the highest power in the QPO feature. 
We used this $\chi^2$ distribution because we averaged in 
frequency by a factor of 128 and averaged 510 individual power spectra. 
This gives a chance probability of $4.8\times 10^{-8}$ for the highest bin 
within the QPO profile, better than a 5$\sigma$ deviation. If one averages the 
two highest bins in the QPO profile then the chance probability per trial drops
to $3.5\times 10^{-13}$. Since we first searched the average power spectra of 
each observation we can be conservative and use as the number of trials the 
total number of powers searched in these power spectra. This gives $5 \times 
512 = 2,560$ trials and then we have a chance probability of 
$9\times 10^{-10}$, better than a $6\sigma$ detection. 

We next modelled the power spectrum using Gaussian profiles for the QPO 
features and a broad band power law component in the 10 - 512 Hz range. 
With this model we found an acceptable fit with $\chi^2 = 485.2$ for 495 
degrees of freedom. Removing the 40 Hz Gaussian feature from the best fit 
model increases $\chi^2$ by $\sim 55$, which further confirms the high 
significance of the 40 Hz QPO. The QPOs at 40 and 67 Hz are well fit by 
Gaussians centered at $\nu_0 = 41.5 \pm 0.4$ and $69.2 \pm 0.15$ Hz, with 
widths of $2.3 \pm 0.47$ and $1.5 \pm 0.14$ Hz respectively. This gives 
coherence values $Q = \nu_0 / \delta\nu_{FWHM} = 19.6$ and 7.7, respectively. 
The average rms amplitudes in the 13 - 27 keV band for the 40 and 67 Hz QPOs 
are $2.4$ and $1.9 \%$. With this model an excess of power remains between 
the two QPOs. The excess can be modelled with a third Gaussian with centroid 
of $56\pm 2$ Hz, width of $13.9\pm 6$ Hz and an rms amplitude of about 2 \%. 
This third Gaussian improves $\chi^2$ by 23.3. An F-test gives a 
probability of $4.8 \times 10^{-5}$, suggesting that the three additional 
parameters are formally required by the data. It is not obvious if this feature
is associated with either the 40 or 67 Hz QPO (ie. a sideband). It is not 
detected in 2 - 12 keV power spectra. We suggest that it may represent a third 
QPO, perhaps a sideband analogous to those identified in neutron star LMXB 
(see Jonker, Mendez \& van der Klis 2000). However, its weakness relative to 
the other QPOs makes such a conclusion tentative until confirmed by subsequent 
observations. Our best fitting model, including all three Gaussian components,
is shown in figure 1 (solid curve).

To investigate the robustness of our result we also computed average power
spectra for several subsets of the data shown in Figure 1. Figure 2 shows a 
pair of power spectra, one computed from the data taken on July 20 and 25, 1997
and the other from the observations on November 17 and 22, 1997. Both the 
40 and 67 Hz QPOs are detected in each of these power spectra. The fact that 
the 40 Hz QPO can be detected at different epochs provides additional 
confidence that it is real. 

We analysed the AO2 data from GRS 1915+105 first because previous publications
had focused on the 67 Hz QPO from AO1 data (proposal 10408, see Morgan, 
Remillard \& Greiner 1997). Having found the 40 Hz QPO in the AO2 data we
went back and carried out a similar analysis on the AO1 data to see if the
40 Hz QPO could be detected. We analysed data from the observatations listed
in Table 2 of Morgan, Remillard \& Greiner (1997). Since not all of these
observations were conducted with the same data modes as used for the AO2
observations we could only directly compare data from the April 29, May 05,
May 14 and June 11, 1996 observations. We computed an average power spectrum
from these observations in the same way as for the AO2 data, again using just
the $> 13$ keV event mode data. We detected the 67 Hz QPO in these data, but 
we did not detect the 40 Hz QPO with an upper limit to the rms amplitude of 
$\sim 0.8 \%$. We found that the centroid frequency of the 67 Hz QPO was
significantly lower in the average AO1 spectrum than for our average 
AO2 spectrum. To investigate this further we plotted our best fit centroids 
from the 5 AO2 observations in which we detected the 67 Hz QPO together with 
the measured centroids and uncertainties from Morgan, Remillard, \& Greiner 
(1997). The results are shown in Figure 3. The AO2 centroids are clearly
systematically higher than the AO1 values by about $\sim 5 \%$. We fit a 
constant frequency model to each set of measurements and found that the AO2 
values are consistent with a constant frequency, whereas the AO1 measurements 
are inconsistent with such a model. We also show with horizontal dashed lines 
the weighted mean for each set of measurements. This clearly demonstrates that 
the 67 Hz QPO can drift in frequency by as much as $5 \%$. 

\section{Discussion and Summary}

There has been a lot of effort expended in recent years in comparing the X-ray
variability properties of neutron stars and black hole binaries. For example,
Psaltis, Belloni, \& van der Klis (1999) investigated correlations between QPO 
frequencies and/or characteristic frequencies of broad band noise components
and suggested that these modulations are caused by similar accretion disk
processes in black hole and neutron star systems. A potential difficulty with 
such studies is the possible ambiguity in associating a particular 
frequency in one class of source with a frequency seen in another. This becomes
more of a concern as more QPO frequencies are identified. GRS 1915+105
is now the second black hole binary to show a pair of high frequency QPOs, and
it would seem a reasonable assumption to conclude that other black holes will 
reveal a similar phenomenology. A crucial question in the 
context of models which attempt to unify X-ray variability of black hole and
neutron star systems is whether or not these QPOs are the analogs of the twin 
kHz QPOs observed in neutron star systems? If this is the case then it would 
appear to rule out models for the kHz variability in neutron stars which 
require the star's solid surface, for example, the sonic point beat-frequency 
model of Miller, Lamb \& Psaltis (1998). If true then it would suggest
that a unified picture of accretion disk modulations may be applicable (see 
for example, van der Klis 1994; Psaltis, Belloni \& van der Klis 1999; 
Psaltis \& Norman 2000; Psaltis 2001).
The phenomenology of QPO pairs in the black holes is not yet well 
developed enough to make a definitive comparison with the better studied
neutron star QPOs, however, their similarities, such as hard energy spectra,
high coherence, and relative frequency spacing, suggest that they 
may indeed be produced by similar processes. At present we suggest this as a 
useful working hypothesis to be tested by further studies. 

A number of different mechanisms have been proposed to explain the high 
frequency QPOs observed in black hole candidates. If the modulation is
caused by orbital motion of material at the inner edge of the disk, then
the required mass for GRS 1915+105 is 33 $M_{\odot}$ if the hole is not 
spinning. For a rapidly spinning black hole the mass could be as high 
as $\sim 200 M_{\odot}$ (Bardeen, Press \& Teukolsky 1972). Nowak et al. 
(1997) suggested the 67 Hz QPO might be explained by a low order `diskoseismic'
$g$-mode. If so, this would require a $\sim 10 M_{\odot}$ non-rotating black
hole or a $\sim 34 M_{\odot}$ maximally rotating hole. It seems unlikely that
a {\it pair} of $g$-modes are responsible because the harmonic spacing would 
be about 4 Hz, and one would therefore require the 40 Hz feature to be a 
higher radial overtone which would not easitly produce an observable amplitude.
Is is conceivable however that the 40 Hz QPO is produced by a 
different oscillation mode. For example, Wagoner (2001) has pointed out that 
the 40 Hz QPO could in principle be identified with a corrugation or $c$-mode 
and the 67 Hz QPO with the $g$-mode. If these identifications are correct one
can obtain constraints on both the mass and spin of the black hole (see
Wagoner 1999; Silbergleit, Wagoner \& Ortega-Rodriquez 2001).

Recent theoretical work has attempted to ascribe the observed QPO frequencies
in neutron star and black hole binaries to General Relativistic frequencies
in the inner accretion disk. In the so called relativistic precession models 
the QPOs have been identified with the Keplerian, the periastron and nodal 
precession frequencies at a characteristic radius in the accretion disk 
(see Stella, Vietri \& Morsink 1999; Psaltis \& Norman 2000). With the 
detection of a second high frequency QPO in GRS 1915+105 we can investigate 
whether the observed QPO frequencies can be
explained self consistently with this model. We show in figure 4 a plot of the
radial epicyclic (upper curves) and nodal precession (diagonal lines) 
frequencies as a function of the Keplerian frequency for a range of black hole 
masses with dimensionless angular momentum $j = 0.12$. 
We also plot the frequencies of QPOs observed from GRS 1915+105 (diamond 
symbols). The 67 and 40 Hz QPOs can be identified with the Keplerian and radial
epicyclic frequency, respectively. A $\sim 0.9$ Hz QPO has been reported by 
Morgan, Remillard \& Grenier (1997) that sometimes appears together with 
the 67 Hz QPO (see Morgan, Remillard \& Grenier 1997). we did not detect such
a QPO during the observations in which we saw the 40 and 67 Hz QPOS. 
For $j \sim 0.12$ such a frequency would be consistent with the nodal 
precession frequency, however, we stress that such an association is tentative 
because the QPOs have not all been seen at the same time. If this model is 
correct it also provides constraints on both the black hole mass and angular 
momentum.  However, we caution that with the plethora of QPO frequencies 
observed from GRS 1915+105, such identifications may not be unique.

We have found a 40 Hz QPO in the hard X-ray flux from the Galactic microquasar 
GRS 1915+105. The source is now the second black hole binary with a pair 
of high frequency QPOs, and it would seem reasonable to assume that this
phenomenology is a property of black holes in general. In particular, the 
similar frequency stability, rms amplitude, and coherence of the 450 and
67 Hz QPOs in GRO J1655-40 and GRS 1915+105 suggest the possibility that the
same physical mechanism is responsible. The pairs of lower frequency peaks
at 300 and 40 Hz, respectively also appear to have similar overall properties.
A plausible scenario is that the same mechanism in the disk produces these
modulations and that differences in the black hole masses accounts for 
the observed frequencies. The Kepler frequency at the ISCO radius scales as 
$M^{-1}$, which would suggest that the black hole in GRS 1915+105 is 
betweem 6 - 7 times more massive than GRO J1655-40. Although a unique
theoretical interpretation is still not agreed upon, further study and in 
particular, independent mass measurements of black holes could test this
assertion and perhaps provide a way to infer black hole masses and spins using
X-ray variability measurements alone. 

\acknowledgements

We thank Craig Markwardt, Jean Swank for many helpful discussions and comments 
on the manuscript. 

\vfill\eject

\section*{References}

\noindent Bandyopadhyay, R. et al. 1998, MNRAS, 295, 623

\noindent Bardeen, J. M., Press, W. H. \& Teukolsky, S. A.
1972, ApJ, 178, 347

\noindent Eikenberry, S. S., Matthews, K., Morgan, E. H., Remillard, R. A., 
\& Nelson, R. W. 1998, ApJ, 494, L61

\noindent Fender, R. P. \& Pooley, G. G. 2000, MNRAS, 318, L1

\noindent Fender, R. P., Pooley, G. G., Brocksopp, C., \& Newell, S. J. 
1997, MNRAS, 290, L65

\noindent Homan, J. et al. 2000, ApJ in press, (astro-ph/0001163)

\noindent Jonker, P. G., Mendez, M. \& van der Klis, M. ApJ, 540, L29

\noindent Leahy, D. A. et al. 1983, ApJ, 266, 160

\noindent Lin, D., Smith, I. A., Liang, E. P., \& B{\"o}ttcher, M. 2000, 
ApJ, 543, L141

\noindent Markwardt, C. B., Swank, J. H. \& Taam, R. E. 1999, ApJ, 513, L37

\noindent Merloni, A., Vietri, M., Stella, L. \& Bini, D. 1999, MNRAS, 304, 155

\noindent Miller, M. C., Lamb, F. K. \& Psaltis, D. 1998, ApJ, 508, 791

\noindent Milsom, J. A., \& Taam, R. E. 1997, MNRAS, 286, 359

\noindent Mirabel, I. F. \& Rodriguez, L. F. 1994, Nature, 371, 46

\noindent Morgan, E. H., Remillard, R. A. \& Greiner, J. 1997, ApJ, 482, 993

\noindent Muno, M. P., Morgan, E. H., \& Remillard, R. A. 1999, ApJ, 527, 321

\noindent Nowak, M. A., Wagoner, R. V. Begelman, M. C. \& Lehr, D. E. 1997, 
ApJ, 477, L91

\noindent Psaltis, D. 2000 ApJ, submitted, (astro-ph/0010316)

\noindent Psaltis, D., Belloni, T. \& van der Klis, M. 1999, ApJ, 520, 262

\noindent Psaltis, D. \& Norman, C. 2000, ApJ, in press (astro-ph/0001391)

\noindent Remillard, R. A., Morgan, E. H., McClintock, J. E., Bailyn, C. D. \& 
Orosz, J. A. 1999, ApJ, 522, 397

\noindent Remillard, R. A. \& Morgan, E. H. 1998, in The Active X-ray Sky, ed. 
L. Scarsi, H. Bradt, P. Giommi, \& F. Fiore (Amsterdam: Elsevier), 316

\noindent Remillard, R. A., Morgan, E. H., McClintock, J. E., Bailyn, C. D. \& 
Orosz, J. A. 1998, in Proc. 18th Texas Symp. on Relativistic Astrophysics, ed. 
A. Olinto, J. Frieman, \& D Schramm (Singapore: World Scientific), 750

\noindent Silbergleit, A. S., Wagoner, R. V., \& Ortega-Rodr{\'i}guez, M. 
2001, ApJ, 548, 335

\noindent Stella, L., Vietri, M. \& Morsink, S. M. 1999, ApJ, 524, L63

\noindent Strohmayer, T. E. 2001, ApJ, 552, L49

\noindent van der Klis, M. 1994, ApJS, 92, 511

\noindent van der Klis, M. 2000, ARAA, in press, (astro-ph/0001167)

\noindent Wagoner, R. V. 2001, personal communication

\noindent Wagoner, R. V. 1999, Phys. Rep., 311, 259


\vfill\eject

\begin{figure*}[htb] 
\centerline{\epsfig{file=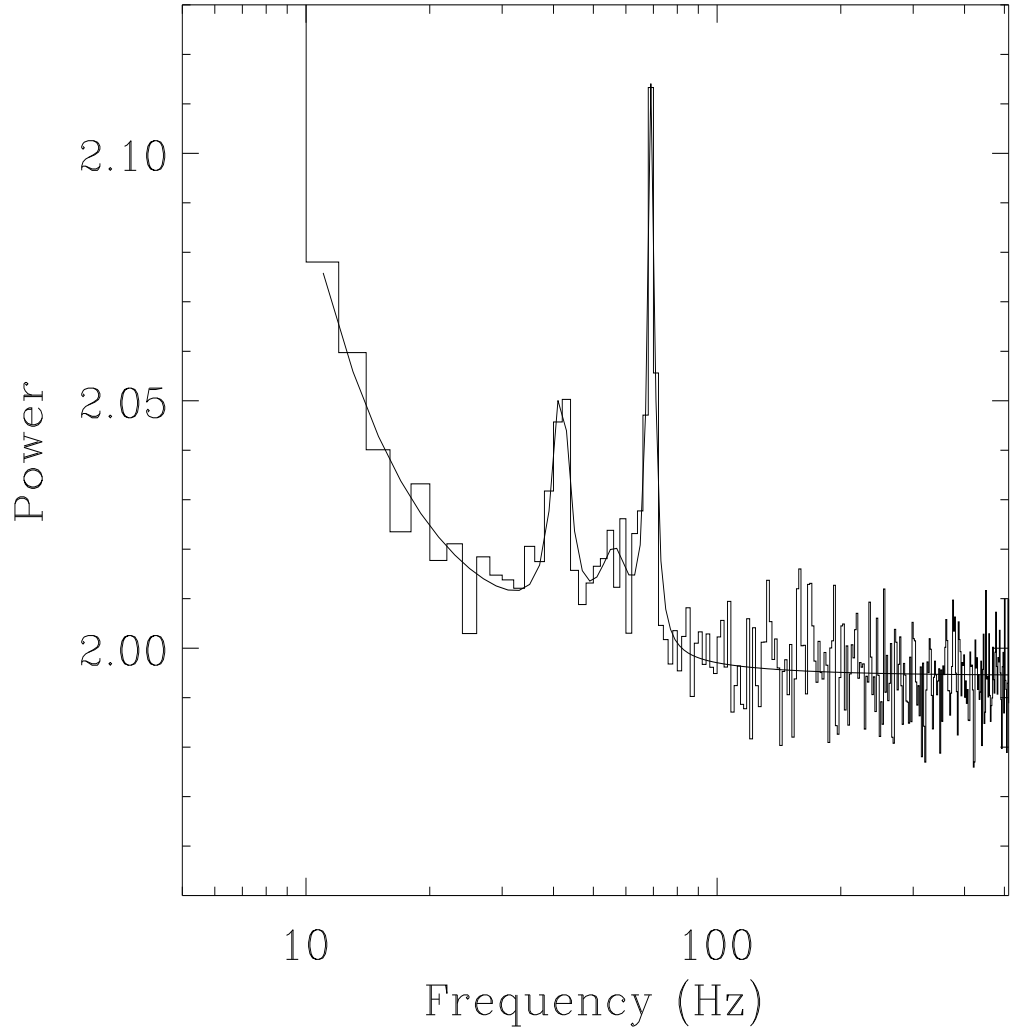,height=6.0in,width=6.0in}}
\vspace{10pt}
\caption{Average power spectrum in the 13 - 27 keV 
energy band for the five observations of GRS 1915+105 listed in Table 1. 
The frequency resolution is 2 Hz and the spectrum has been normalized following
Leahy et al. (1983). The best fitting model is also plotted (solid curve). 
The QPO at 40 Hz is clearly visible.}
\end{figure*}

\vfill\eject

\begin{figure*}[htb] 
\centerline{\epsfig{file=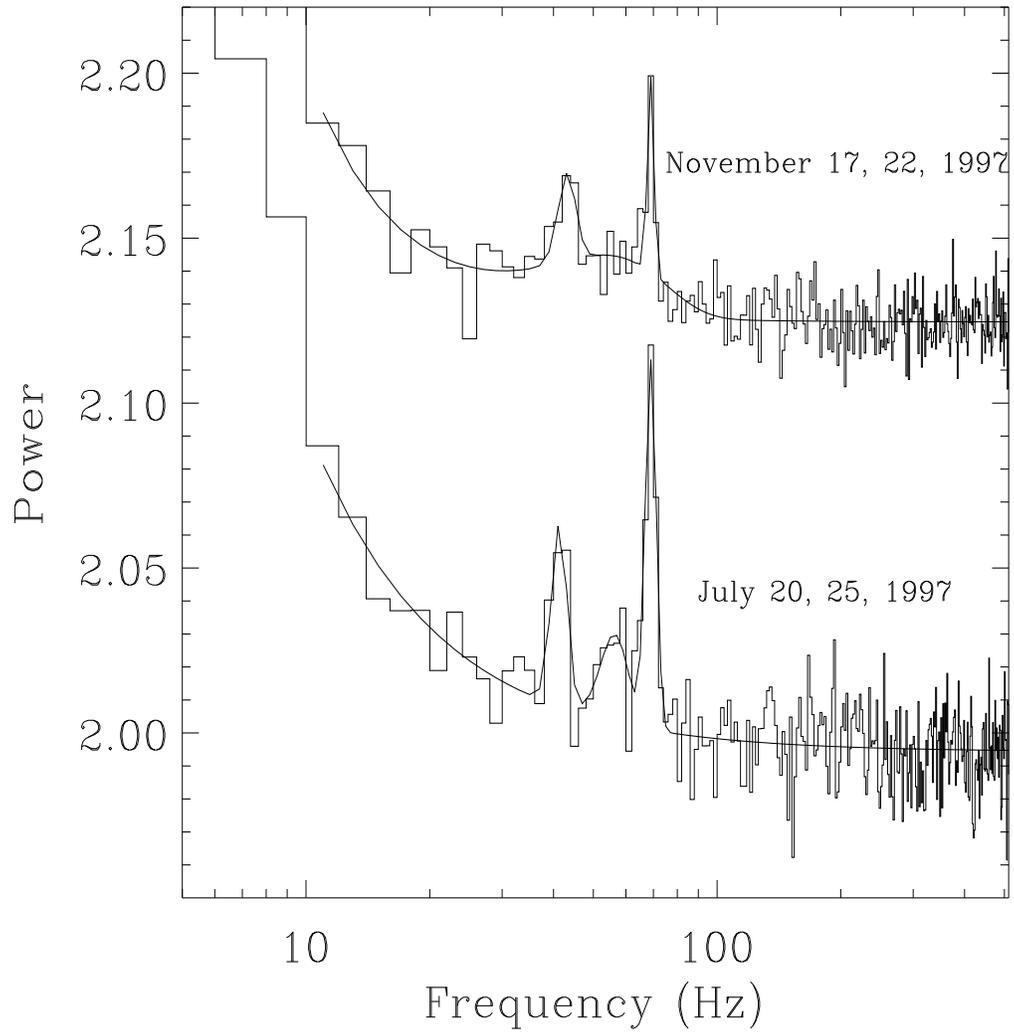,height=6.0in,
width=6.0in}}
\vspace{10pt}
\caption{Average power spectra for the November 17 and 22, 1997 observations 
(top curve), and the July 20, and 25, 1997 data (lower curve). The power 
spectra were calculated in the same manner as in Figure 1. We also show our 
best fitting models (solid curves).}
\end{figure*}

\vfill\eject

\begin{figure*}[htb] 
\centerline{\epsfig{file=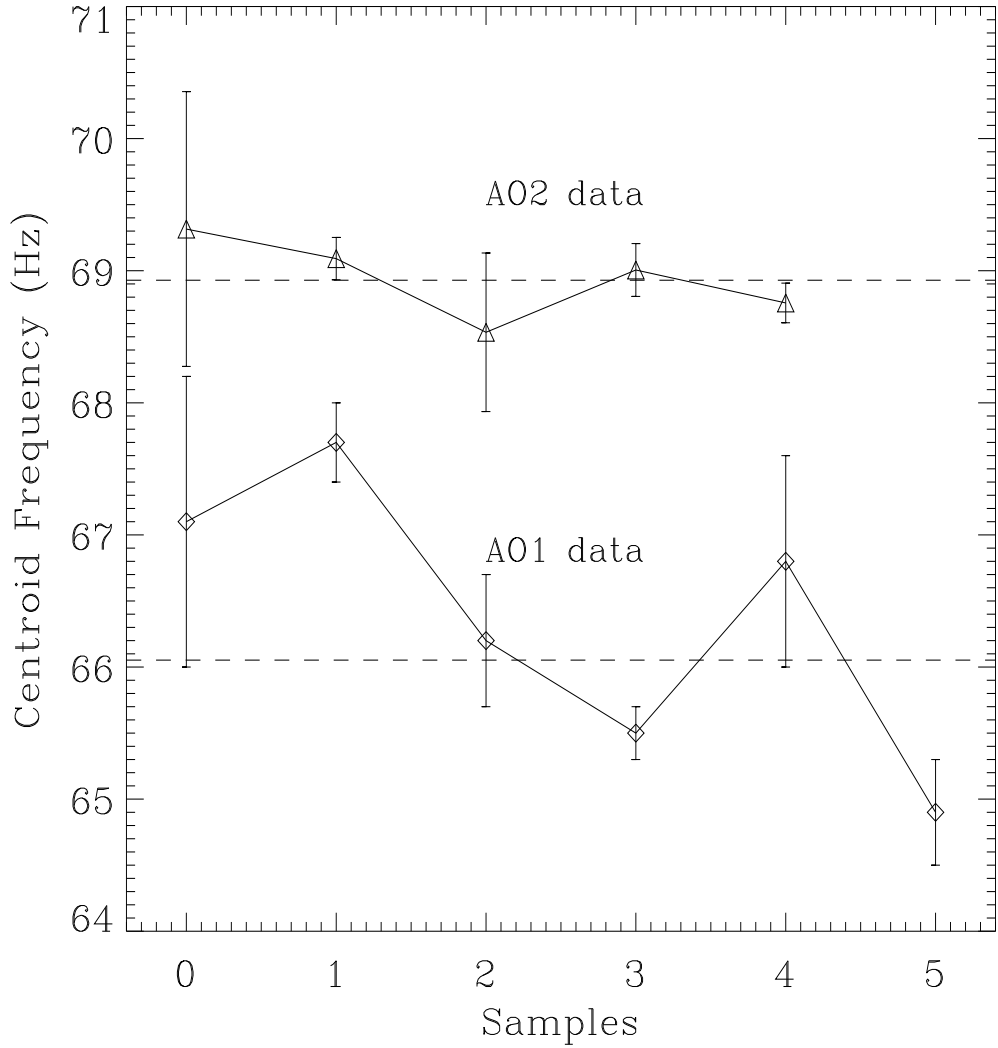,height=6.0in,
width=6.0in}}
\vspace{10pt}
\caption{Comparison of the centroid 
frequencies of the $\sim 67$ Hz QPO in the AO1 (taken from Morgan, 
Remillard, \& Greiner (1997) and AO2 observations. Note that the centroid
frequency is significantly higher in the AO2 observations. The horizontal 
dashed line denotes the weighted mean for each set of measurements. A 
constant frequency is consistent with the AO2 data, but not the AO1 
measurements.}
\end{figure*}

\vfill\eject

\begin{figure*}[htb] 
\centerline{\epsfig{file=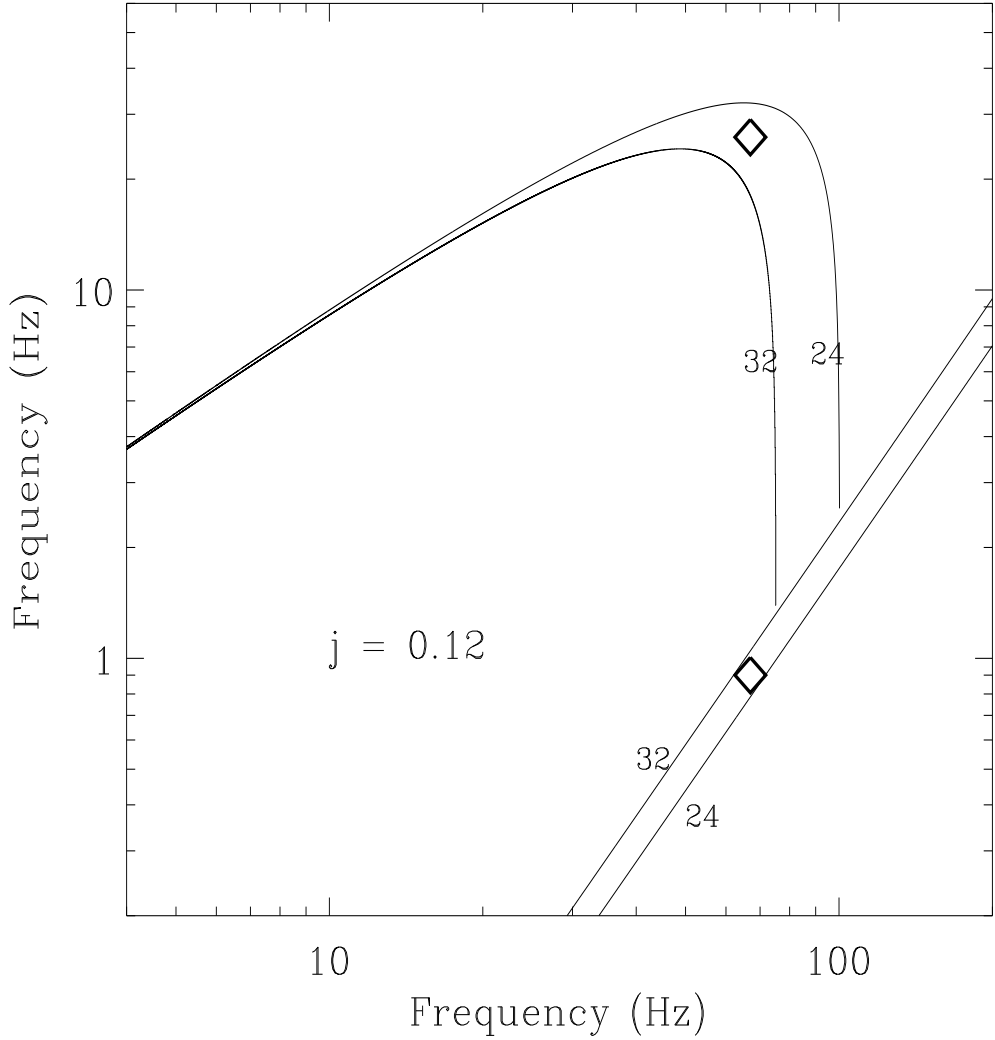,height=6.0in,width=6.0in}}
\vspace{10pt}
\caption{The radial epicyclic frequency (upper
curves) and the nodal precession frequency (lower diagonal lines) plotted
as a function of the Kepler frequency for a Kerr black hole with mass
24 and 32 $M_{\odot}$ and a dimensionless angular momentum, $j = 0.12$.
QPO data are plotted with diamond symbols. Note that the error bars 
are smaller than the symbols. The position of the upper diamond is set by the
40 and 67 Hz QPO. The radial epicyclic frequency is associated with the 
frequency difference between the QPOs. The lower diamond represents the 
$\sim 0.9$ Hz QPO reported by Morgan, Remillard \& Greiner (1997) that was
sometimes present at the same time as the 67 Hz QPO. Since such a QPO was
not detected at the same time as the 40 and 67 Hz QPOs in the data discussed
here its association with the nodal precession frequency is only meant to
be suggestive.}
\end{figure*}

\end{document}